\documentclass{PoS}

\usepackage{relsize}

\title{Phenomenology of $\alpha_s$ at intermediate energy: the quark-hadron duality approach}

\ShortTitle{Phenomenology of $\alpha_s$ at intermediate energy}

\author{\speaker{A. Courtoy}\thanks{This work was funded by the Belgian Fund F.R.S.-FNRS via the contract of Charg\'ee de recherches.}\\
        IFPA, Inst. de Physique, Universit\'e de Li\`ege, Belgium \\
        INFN-LNF, Frascati, Italy\\
        E-mail: \email{aurore.courtoy@ulg.ac.be}}

\abstract{In this contribution to the proceedings, we analyze  the transition from  perturbative and non-perturbative QCD embedded in the coupling constant. In the study of  quark-hadron duality, we suggest that the realization of the latter is related to the inclusion of non-perturbative effects at the level of the coupling constant. The outcome of our analysis is a smooth transition from perturbative to non-pertuperturbative QCD physics, embodied in the running of the coupling constant at intermediate scales. While our approach is purely perturbative, we compare our result to various non-perturbative schemes.}

\FullConference{QCD-TNT-III-From quarks and gluons to hadronic matter: A bridge too far?,\\
		2-6 September, 2013\\
		European Centre for Theoretical Studies in Nuclear Physics and Related Areas (ECT*), Villazzano, Trento (Italy)}

\begin{document}

\section{Introduction}

It is well established  now that the QCD running coupling (effective charge) freezes in the deep infrared, pointing out the breakdown of perturbation theory at in the infrared region. This feature is associated to the transition towards non-perturbative QCD and, therefore, to confinement. In this context, the question of the applicability of quark-hadron duality naturally arises. Quark-hadron duality states that, in specific kinematical regimes, both the perturbative and non-perturbative stages arise almost ubiquitously, in the sense that the non-perturbative description follows the perturbative one.
The knowledge of perturbative QCD can be used to calculate non-perturbative QCD physics observables~\cite{Poggio:1975af}. 
However, when considering perturbative QCD observables at low scales, we implicitly face an interpretation problem. Higher terms in the perturbative expansion of that observable need be taken into account, by definition. Rephrazing, it gives: we are trying to make up for the perturbative to non-perturbative QCD physics transition in the perturbative analysis. We suggest that this phase transition can be fully included in the interpretation of the role of the running coupling constant at the scale of transition instead.

There exists, in Deep Inelastic processes, a dual description between low-energy and high-energy behavior of a same observable, {\it i.e.} the unpolarized structure functions. Bloom and Gilman observed a connection between the structure function $\nu W_2(\nu, Q^2)$ in the nucleon resonance region and that in the deep inelastic continuum~\cite{Bloom:1970xb}: the resonances are not a separate entity but are an intrinsic part of the scaling behavior of $\nu W_2$. The meaning of duality is more intriguing when the equality between resonances and scaling happens at a same scale. It can be understood as a natural continuation of the perturbative to the non-perturbative representation. This context is hence suitable for studying the r\^ole of the running coupling constant at intermediate energies. 

\section{Quark-Hadron Duality in QCD}
\label{sec:lxr}

A quantitative definition of  \emph{global duality} is accomplished by comparing limited intervals  defined according to the experimental data. 
Hence, we analyze the scaling results as a theoretical counterpart, or an output of perturbative QCD, in the same kinematical intervals and at the same scale $Q^2$ as the data for $F_2$. It is easily realized that the ratio,
\begin{eqnarray}
R^{\mbox{\tiny exp/th}}(Q^2)&\equiv&\frac{
\int_{x_{\mbox {\tiny min}}}^{x_{\mbox {\tiny max}}} dx\,
F_2^{\mbox {\tiny exp}} (x, Q^2)
}
{\int_{x_{\mbox{\tiny min}}}^{x_{\mbox{\tiny max}}} dx\,
F_2^{\mbox {\tiny th}} (x, Q^2)
}=1\quad,
\label{eq:ratio_1}
\end{eqnarray}
 if duality is fulfilled.\footnote{In the analysis of Ref.~\cite{Courtoy:2013qca}, we use, for $F_2^{\mbox {\tiny exp}} $, the data from JLab (Hall C, E94110)~\cite{Liang:2004tj} reanalyzed (binning in $Q^2$ and $x$) as explained in~\cite{Monaghan:2012et} as well as the SLAC data~\cite{Whitlow:1991uw}.}

Duality is violated (the ratio~(\ref{eq:ratio_1}) is not $1$) when considering the fully perturbative expression, and is still violated after corrections by the target mass terms.
One possible explanation for the apparent violation of duality is the lack of accuracy in the Parton Distribution Functions (PDF) parametrizations at large-$x$.\footnote{In our analysis, we  use the MSTW08 set at NLO as initial parametrization~\cite{Martin:2009iq}. We have checked that there were no significant discrepancies when using other sets. } Therefore, the behavior of the nucleon structure functions in the 
resonance region needs to be addressed in detail  
in order to be able to discuss 
theoretical predictions in the limit $x \rightarrow 1$. In such a limit, terms containing powers of 
$\ln (1-z)$, $z$ being the longitudinal 
variable in the evolution equations, that are present in 
the Wilson coefficient functions $B_{\mbox{\tiny NS}}^q(z)$ become large and have to be resummed, {\it i.e.} Large-$x$ Resummation (LxR).
Resummation was first introduced by  
linking this issue to the definition of the correct kinematical variable that determines the 
phase space for  real gluon emission
at large $x$. This was found to be $\widetilde{W}^2 = Q^2(1-z)/z$, 
instead of $Q^2$~\cite{Amati:1980ch}.
As a result, the argument of the strong coupling constant becomes $z$-dependent~\cite{Roberts:1999gb},
\begin{equation} 
\alpha_s(Q^2) \rightarrow \alpha_s\left(Q^2 \frac{(1-z)}{z}\right)\quad.
\end{equation}

In this procedure, however, an ambiguity is introduced, related to the need of continuing 
the value of $\alpha_s$  
for low values of its argument, {\it i.e.} for $z \rightarrow 1$. 
In Ref.~\cite{Courtoy:2013qca}, we have reinterpretated  $\alpha_s$ for values of the scale in the infrared region. 
To do so, we investigated the effect induced by changing the argument of $\alpha_s$ on the behavior of the $\ln(1-z)$-terms in the convolution with the coefficient function $B_{\mbox{\tiny NS}}$: %
\begin{eqnarray}
F_2^{NS} (x, Q^2) 
&=&x q(x,Q^2)+ \frac{\alpha_s}{4\pi} \mathlarger{\sum}_q \mathlarger{\int}_x^1 dz \, B_{\mbox{\tiny NS}}^q(z) \, \frac{x}{z}\, q\left(\frac{x}{z},Q^2\right)\quad,
\label{eq:evo}
\end{eqnarray}
We resum those terms as
\begin{eqnarray}
\ln(1-z)&=&\frac{1}{\alpha_{s, \mbox{\tiny LO}}(Q^2) }\int^{Q^2} d\ln Q^2\, \left[\alpha_{s, \mbox{\tiny LO}}(Q^2 (1-z)) -\alpha_{s, \mbox{\tiny LO}}(Q^2) \right]
\equiv \ln_{\mbox{\tiny LxR}}\quad,
\end{eqnarray}
 including the complete $z$ dependence of $\alpha_{s, \mbox{\tiny LO}}(\tilde W^2)$ to all logarithms.
Using the `resummed'  $F_2^{\mbox{\tiny theo}} $ in Eq.~(\ref{eq:ratio_1}), the ratio $R$ decreases substantially, even reaching values lower than 1. It is a consequence of the change of the argument of the running coupling constant. At fixed $Q^2$, under integration over $x<z<1$, the scale $Q^2\times(1-z)/z$ is shifted  and can reach low values, where the running of the coupling constant starts blowing up. At that stage, our analysis requires non-perturbative information.

In the light of quark-hadron duality, it is necessary to prevent the evolution from enhancing the scaling contribution over the resonances.  We define the limit from which non-perturbative effects have to be accounted for by setting a maximum value for the longitudinal momentum fraction, $z_{max}$. Two distinct regions can be studied: the ``running" behavior  in $x<z<z_{max}$ and the ``steady" behavior $z_{max}<z<1$. 
Our definition of the maximum value for the argument of the running coupling follows from the realization of duality in the resonance region. The value $z_{max}$ is reached at
\begin{eqnarray}
R^{\mbox{\tiny exp/th}}(z_{\mbox{\tiny max}}, Q^2)
&=&\frac{
 \mathlarger{\int}_{x_{\mbox {\tiny min}}}^{x_{\mbox {\tiny max}}} dx\,
F_2^{\mbox {\tiny exp}} (x, Q^2)
}
{ \mathlarger{\int}_{x_{\mbox{\tiny min}}}^{x_{\mbox{\tiny max}}} dx\,
F_2^{NS, \mbox{\tiny Resum}} (x, z_{\mbox{\tiny max}}, Q^2)
}
= \frac{I^{\mbox{\tiny exp}}}{I^{ \mbox{\tiny Resum}}}= 1\quad.
\label{eq:ratio_max}
\end{eqnarray}
%

\section{The Running Coupling Constant}
\label{sec:alpha}

\begin{figure}[h]
\centering
\includegraphics[scale= 1.]{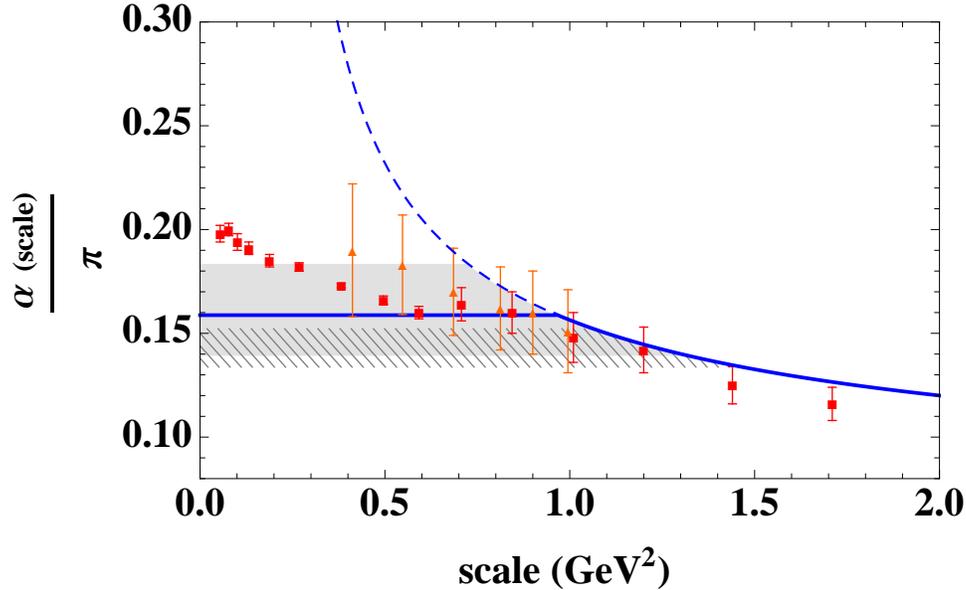}
\caption{Extraction of $\alpha_s$.  See text. 
\label{extract}}
\end{figure}

The direct consequence of the previous Section is that duality is realized, within our assumptions, by allowing $\alpha_s$  to run from a minimal  scale only. From that minimal scale downward, the coupling constant does not run, it is frozen. This feature is illustrated on Fig.~\ref{extract}. We show the behavior of $\alpha_{s, \mbox{\tiny NLO}}$(scale) in the $\overline{\mbox{MS}}$ scheme and for the same value of $\Lambda_{\overline{\mbox{\scriptsize MS}}, \mbox{\scriptsize MSTW}}^{\mbox{\scriptsize NLO}}=0.402$ GeV used throughout our analysis. 
The theoretical errorband correspond to the extreme values of
\begin{equation}
\alpha_{s, \mbox{\tiny NLO}}\left(Q_i^2\frac{(1-z_{max, i})}{z_{max, i }}\right) \qquad,
\end{equation}
$i$ corresponds to the data points.
Of course, we expect the transition from non-perturbative to perturbative to occur at one unique scale. The discrepancy between the 10 values we have obtained has to be understood as the resulting error propagation. The grey area represents the approximate frozen value of the coupling constant,
\begin{equation}
0.13\leq \frac{\alpha_{s, \mbox{\tiny NLO}} (\mbox{scale}\to 0 \mbox{GeV}^2)}{\pi} \leq 0.18 \quad.
\end{equation}
The solid blue curve represents the (mean value of the) coupling constant obtained from our analysis using inclusive electron scattering data at large $x$.  The blue dashed curve represents the exact NLO solution for the running coupling constant in $\overline{\mbox{MS}}$ scheme.  The grey area represents the region where the freezing occurs for JLab data, while the hatched area corresponds the freezing region determined from SLAC data. This  error band represents  the theoretical uncertainty in our analysis. 

%
In the figure we also report  values from the extraction using polarized $eP$ scattering data in Ref.~\cite{Deur:2005cf,Deur:2008rf,alexandre}. These values represent the  first extraction of an effective coupling in the IR region that was obtained by analyzing the data  
relevant for the study of the GDH sum rule. To extract the coupling constant,  the $\overline{\mbox{MS}}$ expression of the Bjorken sum rule up to the 5th order in alpha (calculated in the $\overline{\mbox{MS}}$ scheme) was used. The red squares correspond to $\alpha_s$ extracted from Hall B CLAS EG1b, with statistical uncertainties; the  orange triangles corresponds to Hall A E94010 / CLAS EG1a
data, the uncertainty here contains both statistics and systematics.
The agreement with our analysis, which is totally independent, is impressive.
We notice, and it is probably one of the most important result of our analysis, that the transition from perturbative to non-perturbative QCD seems to occur around $1$ GeV$^2$.

\begin{figure}[tb]
\centering
\includegraphics[scale= 1.]{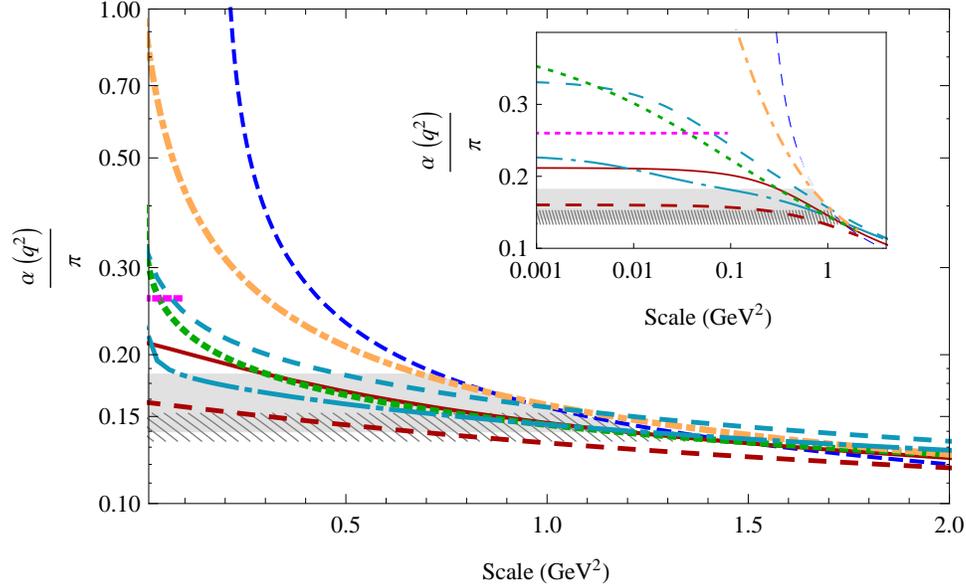}
\caption{ Comparison of the effective coupling constant. See text. }
\label{comparisons}
\end{figure}

At that stage, a comparison with fully non-perturbative effective charges and ``modified pQCD" is noteworthy. It is shown in Fig.~\ref{comparisons}.
The grey areas are as in Fig.~\ref{extract} with $\Lambda_{\overline{\mbox{\scriptsize MS}}, \mbox{\scriptsize MSTW}}^{\mbox{\scriptsize NLO}}=0.402$ GeV ; the dashed blue curve is the exact NLO solution with the same $\Lambda$. 
The dotted-dashed orange curve corresponds to the result of Ref.~\cite{Fischer:2003rp}, using the version $(b)$ of their
fit  with $a=b=1$. The latter analysis was performed in the MOM renormalization scheme. Though the $\beta$ function does not depend on the scheme up to 2 loops, the definition of $\Lambda$ varies from scheme to scheme. The comparison of the results is made possible using the relation,~\cite{Boucaud:1998bq}
\begin{equation}
\Lambda_{\overline{\mbox{\scriptsize MS}}}=\frac{\Lambda_{\mbox{\scriptsize MOM}}}{3.334}\quad,
\end{equation}
leading to the value of $\Lambda_{\overline{\mbox{\scriptsize MS}}}^{\mbox{\scriptsize Ref.~\cite{Fischer:2003rp}}}=(0.71/3.334) $ GeV$\sim 0.21$GeV. The value of $\alpha(0)$ is fixed to $8.915/N_c$.
The red curves are variations of the effective charge of Ref.~\cite{Cornwall:1981zr},
\begin{eqnarray}
\frac{\alpha(Q^2)}{4\pi} &=& \left[\beta_0 \ln \left(\frac{Q^2 +\rho m^2(Q^2)}{\Lambda^2}\right)\right]^{-1}\\
\label{eq:cornwall}
\mbox{with} \nonumber&&\\
m^2 (Q^2)&=& m^2_0\left[\ln\left(\frac{Q^2 + \rho m_0^2}{\Lambda^2}\right)
\bigg/\ln\left(\frac{\rho m_0^2}{\Lambda^2}\right)\right]^{-1 -\gamma}\quad,\nonumber
\end{eqnarray}
where $(m_0^2, \rho, \Lambda)$ are parameters to be fixed. The solid red curve corresponds to the set $(m_0^2=0.3 $GeV$^2, \rho=1.7, \Lambda=0.25$GeV$)$,  the dashed red curve to $(m_0^2=0.5 $GeV$^2, \rho=2., \Lambda=0.25$GeV$)$.
This result is also obtained in the MOM scheme, the value of $\Lambda$ turns out to be similar in both Fischer {\it et al.} and Cornwall's approaches. The cyan curves correspond to two scenarios of the effective charges of Ref.~\cite{Aguilar:2009nf}. Their numerical solution is fitted by a functional form similar to Eq.~(\ref{eq:cornwall}). The 2 sets of parameters, corresponding to $m_0=500$MeV (dashed-dotted curve) and $600$ MeV (medium dashed curve), are then driven by the shape of the numerical solution. They are plotted here with the same $\Lambda_{\mbox{\tiny MOM}}^{n_f=0}=300$ MeV as in the publication, but for $n_f=3$ for sake of comparison. Further investigation on comparison of  schemes is needed.
The short dashed green curves corresponds to Shirkov's analytic perturbative QCD to LO~\cite{Shirkov:1997wi} with $\Lambda_{\overline{\mbox{\scriptsize MS}}, \mbox{\scriptsize MSTW}}^{\mbox{\scriptsize LO}}$. The value of $\alpha(0)$ is fixed to $4\pi/\beta_0$.  Finally, the pink curve is the freezing value of Ref.~\cite{Mattingly:1992ud}. 

\section{Conclusions}

We report an interesting observation that  the values of the coupling from different measurements/observables 
namely the GDH sum rule~\cite{alexandre}, and the large-$x$-DIS/resonance region based extractions, 
are in very good agreement.  The extraction from the GDH sum rule, in a different (observable) scheme~\cite{Deur:2005cf,Deur:2008rf}, turns out to be in agreement with the prediction from AdS/CFT~\cite{Brodsky:2010ur}. A comparison of our result in the $\overline{\mbox{\scriptsize MS}}$ scheme requires the extension to observable dependence~\cite{Grunberg:1980ja} from  scheme dependence. It will be studied in a future publication.

We have also compared our extraction to  non-perturbative approaches. Notice that the extracted value for $\alpha_s(Q^2<1$GeV$^2)$ is only constrained by the integral in the resummed version of Eq.~(\ref{eq:evo}): no conclusion can be drawn on its value at $Q^2=0$GeV$^2$.   While it is not possible to conclude on the value of $\alpha_s(0)$, we notice that it is possible to find  sets of parameters for which the transition from perturbative to non-perturbative QCD occurs around $1$ GeV$^2$.


\end{document}